# Impact of Rushing attack on Multicast in Mobile Ad Hoc Network

## V. PALANISAMY[1], P.ANNADURAI[2],

1  Reader and Head (i/c), Department of Computer Science & Engineering,
Alagappa University, Karaikudi, Tamilnadu ,India
Email: vpazhanisamy@yahoo.co.in
2 Lecturer in Computer Science, Kanchi Mamunivar Centre for
Post Graduate Studies (Autonomous) , Lawspet, Puducherry, India.
Email: annadurai_aps70@yahoo.co.in

*Abstract*— A mobile ad hoc network (MANETs) is a self-organizing system of mobile nodes that communicate with each other via wireless links with no fixed infrastructure or centralized administration such as base station or access points. Nodes in a MANETs operate both as host as well as routers to forward packets for each other in a multi-hop fashion.  For many applications in wireless networks, multicasting is an important and frequent communication service. By multicasting, since a single message can be delivered to multiple receivers simultaneously.  It greatly reduces the transmission cost when sending the same packet to multiple recipients.

The security issue of MANETs in group communications is even more challenging because of involvement of multiple senders and multiple receivers.  At that time of multicasting, mobile ad hoc network are unprotected by the attacks of malicious nodes because of vulnerabilities of routing protocols. Some of the attacks are Rushing attack, Blackhole attack, Sybil attack, Neighbor attack and Jellyfish attack.

This paper is based on Rushing attack. In Rushing attack, the attacker exploits the duplicate suppression mechanism by quickly forwarding route discovery packets in order to gain access to the forwarding group and this will affect the Average Attack Success Rate.

In this paper, the goal is to measure the impact of Rushing attack and their node positions which affect the performance metrics of Average Attack Success Rate with respect to three scenarios: near sender, near receiver and anywhere within the network. The performance of the Attack Success Rate with respect to above three scenarios is also compared.

## I. INTRODUCTION

A mobile ad hoc network is a self-organizing system of mobile nodes that communicate with each other via wireless links with no infrastructure or centralized administration such as base stations or access points. Nodes in MANET operate both as hosts as well as routers to forward packets for each other in a multi-hop fashion. MANETSs are suitable for applications in which no infrastructure exists such as military battlefield, emergency rescue, vehicular communications and mining operations.

In these applications, communication and collaboration among a given group of nodes are necessary. Instead of using multiple unicast transmissions, it is advantageous to use  multicast in order to save network bandwidth and resources,  since a single message can be delivered to multiple receivers simultaneously. Existing multicast routing protocols in MANETs can be classified into two categories: tree based and mesh-based.  In a multicast routing tree, there is usually only one single path between a sender and a receiver, while in a routing mesh, there may be multiple paths between each sender receiver pair. Routing meshes are thus suitable than routing trees for systems with frequently changing topology such as MANETs due to availability of multiple paths between a source and a destination. Example tree-based multicast routing protocols are MAODV, AMRIS, BEMRP, and ADMR.  Typically  mesh-based  multicast  routing protocols are ODMRP, FGMP, CAMP , DCMP , and NSMP [2].

Among all the research issues, security is an essential requirement in MANET environments. Compared to wired networks, MANETs are more vulnerable to security attacks due to lack of trusted centralized authority, lack of trust relationships between mobile nodes, easy eavesdropping because of shared wireless medium, dynamic network topology, low bandwidth, and battery and memory constraints of mobile devices. The security issue of MANETs in group communications is even more challenging because of the involvement of multiple senders and multiple receivers. Although several types of security attacks in MANETs have been studied in the literature, the focus of earlier research is on unicast (point to point) applications. The impacts of security attacks on multicast in MANETs have not yet been explored [3].

In this paper, we present a simulation-based study of the effects of Rushing attack on multicast in MANETs. We consider the most common types of attacks, namely rushing attack, blackhole attack, neighbor attack and jellyfish attack.

### A. Goal

The goal of this paper is to impact of rushing attack on mesh-based multicast in MANETs. *The rushing attack*, that acts as an effective denial-of-service attack against all currently proposed on-demand ad hoc network routing protocols, including protocols that were designed to be secure. [2]

In this work, to simulate three scenarios: The attacker node is place at near sender, the attacker node is place at near receiver. The attacker node is place anywhere within the MANETs. Based on above scenarios, to simulate how the Rushing attack affects the network performance.

### B. Reading Roadmap

This paper starts with this section, which gives a brief introduction, and goal of this paper. **Section 2** describes preliminaries for multicast attacks in MANETs. The Improved model scheme Impact of Rushing Attack on Multicast in Mobile Ad hoc Networks (IRAMA) is presented in **Section 3**. **In Section 4**, we discuss the experimental results and discussion. Finally, conclusions are given in **Section 5.**

## II. MULTICAST AND ITS ATTACKS IN MOBILE AD HOC NETWORK

### A. Introduction

A mobile ad hoc network (MANETs) is a self-organizing system of mobile nodes that communicate with each other via wireless links with no fixed infrastructure or centralized administration such as base station or access points. Nodes in a MANETs operate both as host as well as routers to forward packets for each other in a multi-hop fashion. For many applications in wireless networks, multicasting is an important and frequent communication service. By multicasting, since a single message can be delivered to multiple receivers simultaneously. It greatly reduces the transmission cost when sending the same packet to multiple recipients [4, 5].

Multicast is communication between a single sender and multiple receivers on a network. Otherwise it transmits a single message to a select group of recipients. Multicast is used, for example, in streaming video, in which many megabytes of data are sent over the network. Single packets copied by the network and sent to a specific subset of network addresses. These addresses are specified in the Destination Address. Protocol to allow point to multipoint efficient distribution of packets, frequently used in access grid applications. It greatly reduces the transmission cost when sending the same packet to multiple recipients. The option to multicast was made possible by digital technology to allow each digital broadcast station to split its bit stream into 2, 3, 4 or more individual channels of programming and/or data services.

Instead of using multiple unicast transmissions, it is advantageous to use multicast in order to save bandwidth and resources. Since a single message can be delivered to multiple receivers simultaneously. Multicast data may still be delivered to the destination on alternative paths even when the route breaks. It is typically used to refer to IP multicast which is often employed for streaming media and At the Data Link Layer, *multicast* describes one-to-many distribution such as Ethernet multicast addressing, Asynchronous Transfer Mode (ATM) point-to-multipoint virtual circuits or Infiniband multicast. Teleconferencing and videoconferencing also use multicasting, but require more robust protocols and networks. Standards are being developed to support multicasting over a TCP/IP network such as the Internet. These standards, IP Multicast and Mbone, will allow users to easily join multicast groups. [6]

### B. Attack against ad hoc network

While a wireless network is more versatile than a wired one, it is also more vulnerable to attacks. This is due to the very nature of radio transmissions, which are made on the air. On a wired network, an intruder would need to break into a machine of the network or to physically wiretap a cable. On a wireless network, an adversary is able to eavesdrop on all messages within the emission area, by operating in promiscuous mode and using a packet sniffer (and possibly a directional antenna). Furthermore, due to the limitations of the medium, communications can easily be perturbed; the intruder can perform this attack by keeping the medium busy sending its own messages, or just by jamming communications with noise. [1]

Security has become a primary concern to provide protected communication between mobile nodes in a hostile environment. Unlike wireline networks, the unique characteristics of mobile ad hoc networks pose a number of non-trivial challenges to the security design. Providing security support for mobile ad-hoc networks is challenging for several reasons: (a) wireless networks are susceptible to attacks ranging from passive eavesdropping to active interfering, occasional break-ins by adversaries (b) mobile users demand "anywhere, anytime" services; (c) a scalable solution is needed for a large-scale mobile network (d) Dynamic topology (e) infrastructure less (f) Peer –to-peer network (g) Lack of centralized authority [17].

### C Attacks on Multicast

Multicast conserves network bandwidth by sending a single stream of data to multiple receivers. Packets are duplicated only at branch points. The security issue of MANETs in group communications is even more challenging because of involvement of multiple senders and multiple receivers. Some different types of multicast attacks are Rushing attack, Balckhole attack, Neighbor attack, and Jellyfish attack.

*D. Rushing Attack*

A rushing attacker exploits this duplicate suppression mechanism by quickly forwarding route discovery packets in order to gain access to the forwarding group. [8]

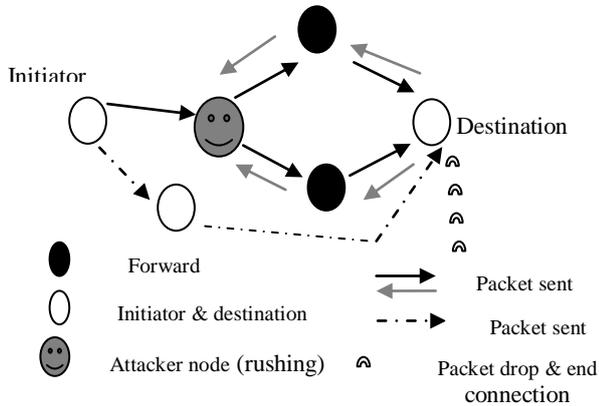

Figure 1 Rushing Attack

- Goal: to invade into routing paths
- Target: multicast routing protocols that use a *duplicate suppression mechanism* in order to reduce routing overheads.
- Method: quickly forwards route discovery (control) packets by skipping processing or routing steps. Rushing attack otherwise, falsely sending malicious control messages and then forwards the packet fastly than clear node reachable.

*E. BlackHole Attack*

An attacker can drop received routing messages, instead of relaying them as the protocol requires, in order reducing the quantity of routing information available to the other nodes.

This is called *black hole attack*, and is a "passive" and a simple way to perform a Denial of Service. The attack can be done selectively (drop routing packets for a specified destination, a packet every *n* packets, a packet every *t* seconds, or a randomly selected portion of the packets) or in bulk (drop all packets), and may have the effect of making the destination node unreachable or downgrade communications in the network.

*Message Tampering*

An attacker can also modify the messages originating from other nodes before relaying them, if a mechanism for message integrity (i.e. a digest of the payload) is not utilized.

A packet drop attack or black hole attack is a type of denial-of-service attack accomplished by dropping packets. Black holes refer to places in the network where incoming traffic is silently discarded (or "dropped"), without informing the source that the data did not reach its intended recipients [8,9].

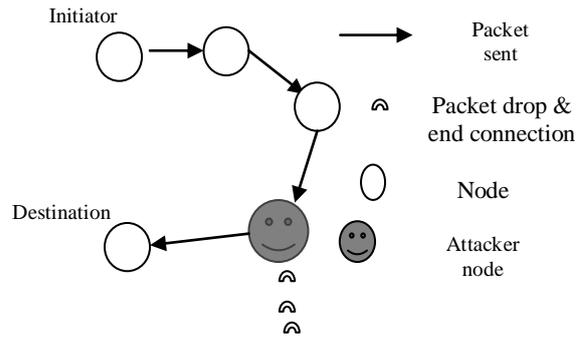

Figure 2 (a). Black Hole Attack (Drop all packets)

- Goal: to damage the packet delivery ratio
- Target: all multicast protocols
- Method: an attacker
  o First invades into forwarding group (e.g., by using rushing attack),
- Then drops some or all data packets instead of forwarding them

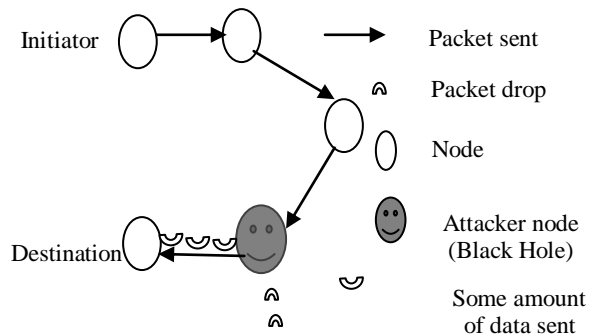

Figure 2(b) Black Hole attack (small amt of data only drop)

Black Hole attacks effects the packet delivery and to reduce the routing information available to the other nodes

Causes:

- It down grade the communication

- Effects of making the destination node reachable

*F. Neighbor Attack*

Upon receiving a packet, an intermediate node records its Id in the packet before forwarding the packet to the next node. An attacker, however, simply forwards the packet without recording its Id in the packet to make two nodes that are not within the communication range of each other believe that they are neighbors (i.e., one-hop away from each other ), resulting in a disrupted route.

### G.  Jelly Fish Attack

A jellyfish attacker first needs to intrude into the multicast forwarding group. It then delays data packets unnecessarily for some amount of time before forwarding them. This results in significantly high end-to-end delay and thus degrades the performance of real applications.

*Causes:*

Increase end –end delay.

### H.  Sybil Attack

Sybil attack manifests itself by allowing the malicious parties to compromise the network by generating and controlling large numbers of shadow identities. The fact is that each radio represents a single individual. However the broadcast nature of radio allows a single node to pretend to be many nodes simultaneously by using many different addresses while transmitting. The off-shoot of this Sybil attack is analyzed using Packet Delivery Ratio (PDR) as the performance metric. Theoretical based graphs are simulated to study the influence of Sybil attack in PDR [18].

Malicious user obtaining multiple fake identifies and pretends to be multiple distinct node in the system malicious node control the decision of the system [8]. The Sybil attack can be categorized into sub categories: presentation of multiple identities simultaneously and presentation of multiple identities exclusively.

The concept of the identifiers exists at different levels and because an identifier only guarantees the uniqueness at the intended level only. Sybil attack can be perpetrated from network layer and application layer where the respective identifiers are IP address and Node ID. Sybil attack can be manifested either by creating new identities or duplicating other identities by disabling them after launching a DoS attack. This mechanism can be either a localized or globalized one depending on the severity of the attack felt by neighboring nodes. Sybil attack can defeat the objectives of distributed environment like fair resource allocation, voting, routing mechanism, distributed storage, misbehavior detection etc.

### III.  IMPROVED MODEL (IMPACT OF RUSHING ATTACK ON MULTICAST IN MOBILE AD HOC NETWORK

### A.  Related Work

In this related work to measure a simulation-based study of the effects of Rushing attacks on multicast in MANETs. A Rushing attacker first needs to invade into the multicast forwarding group in order to capture data packets of the multisession. If then they quickly forward the data packets to the next node on the routing path. This type of attack often results in very low Average Attack Success Rate [15].

### B.  Rushing Attack and its Impacts in Ad hoc Networks

Multicast is communication between a single sender and multiple receivers on a network. Otherwise it transmits a single message to a select group of recipients. On a wireless network, an adversary is able to eavesdrop on all messages within the emission area, by operating in promiscuous mode and using a packet sniffer (and possibly a directional antenna). Furthermore, due to the limitations of the medium, communications can easily be perturbed; MANETS are more vulnerable to attacks than wired networks due to open medium, dynamically changing network topology, cooperative algorithms, lack of centralized monitoring and lack of clear line of defense [10].

Typically, multicast on-demand routing protocols state that nodes must forward only the first received Route Request from each route discovery; all further received Route requests are ignored. This is done in order to reduce cluttering. The attack consists, for the adversary, in quickly forwarding its Route Request messages when a route discovery is initiated. If the Route Requests that first reach the target's neighbors are those of the attacker, then any discovered route includes the attacker. *The rushing attack*, that acts as an effective denial-of-service attack against all currently proposed on-demand ad hoc network routing protocols, including protocols that were designed to be secure. [14] In this work, to simulate three scenarios:

- The attacker node is place at near sender
- The attacker node is place at near receiver.
- The attacker node is place anywhere within the network.

Based on above scenarios, to simulate how the Rushing attack affects the network performance.

### C.  Rushing Attack Formation

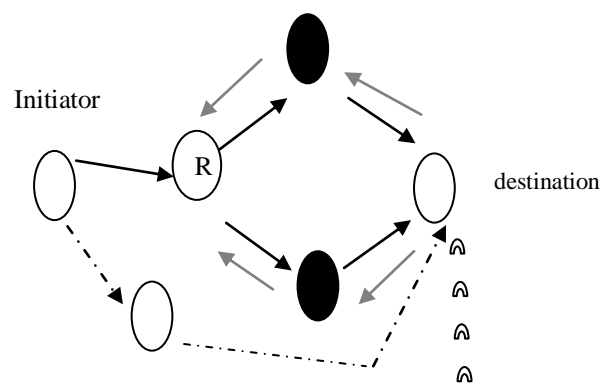

Figure 3 Rushing attack Formation

*Algorithm for Rushing Attack Formation*

Step1: Set of N number of nodes are created.
Step2: Create a connection between nodes.
Step3: Rushing node invaded into the forward multicast
       group.
Step4: Send the packet to the particular groups
Step5: At mean time attacker node tap all the packets.
Step6: The packets in the attacker node are then quickly
       forwarded to the next upcoming node.
Step7: The data packets from the legitimate node reaches
       the destination late and so it is dropped as
       duplicate packet.
Step8: Rushing node in the multicast grouping, affect the
       Avg Attack Success Rate.

*C       Rushing Attack Based on Three scenarios*

*i. Rushing attack at near sender*

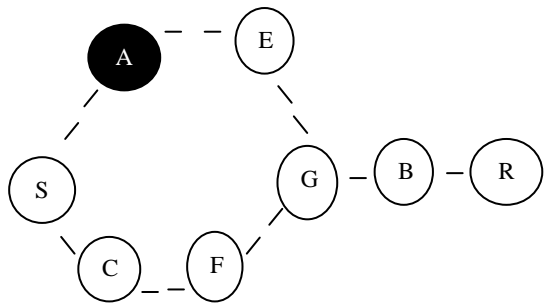

Figure 4. Rushing Node at near Sender

In this figure 4 node S sends the packet to the
destination node R. The attacker node A is placed at near
sender. The data packets from the sender are forwarded to
both the node A and C at the same time. The attacker
nodes quickly forward the data packet to node E than the
node C. The attacker node forwards the packet to node E
then to G and B node. Finally Receiver R receives the
data packets that are forwarded by attacker node. The
performance of Attack Success Rate with respect to this
scenario is calculated.

*Algorithm for near sender*

Step 1**:** Create a set of n number of nodes
Step2: Create a connection between the nodes
Step3: Invade the attacker node at near sender
Step4: Sender sends the packet through specified path.
Step5: Other forward nodes, forward the packet to the
       next node.
Step6: The attacker node taps all the packets.
Step7: The attacker node quickly forwards the packets to
       the next node that are closest to the receiver
Step8: The data packets are then finally reaches the
       destination node.

*ii. Rushing attack at near receiver*

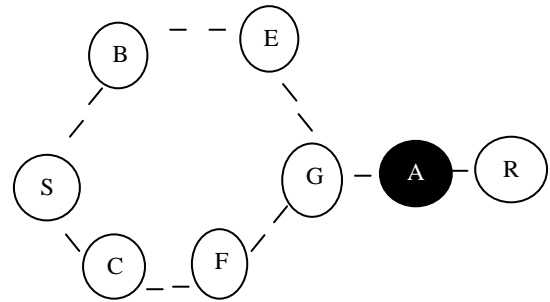

Figure 5 Rushing Node at near Receiving

In this figure 5 node S sends the packet to the
destination node R. The attacker node A is placed at near
receiver. The sender node forwards the data packets to
both the node B and C at the same time. The data packet
can pass through either B, E and G nodes or C, F and G
nodes. When the data packet reaches the attacker node A,
it quickly forwards the data packet to node R. The
performance of Attack Success Rate with respect to this
scenario is calculated.

*Algorithm for near receiver*

Step 1: Create a set of n number of nodes.
Step2: Create a connection between the nodes.
Step3: Invade the attacker node at near receiver.
Step4: Sender send the packets through specified path.
Step5: Other forward nodes, forward the packet to the
       next node.
Step 6: Attacker node tap all the packets through the
       specified path.
Step7: The attacker node then quickly forwards the
       packets.
Step8: Intermediate node forwards the packets to the
       destination node .

*iii. Rushing attack at anywhere within the network:*

In this figure 5 node S sends the packet to the
destination node R. The attacker node A is placed
anywhere within the network. The data packet from the
sender is forwarded to the nodes B and C. The data
packet is then forwarded through the nodes B and E. But
the data packet passed through the node C and then to
attacker node A which quickly forwards the data packet
to the node G than from the node E. The data packet is
then finally reaches the receiver node R through node F.
The performance of Attack Success Rate with respect to
this scenario is calculated.

*Algorithm for anywhere within network*

Step 1: Create a set of n number of nodes
Step2: Create a connection between the nodes
Step3: Invade the attacker node at anywhere within the
       network

Step4: Sender send the packet through specified path.
Step5: Other forward nodes, forward the packet to the next node.
Step6: The attacker nodes tap the entire packet.
Step7: The attacker node then quickly forwards the packets.
Step8: The intermediate node forwards packet to the next node until it reaches the destination.

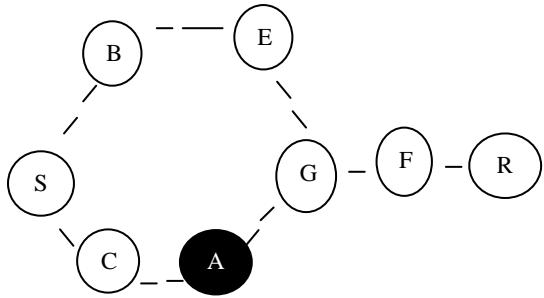

Figure 6 Rushing Node at anywhere within the network.

## IV. EXPERIMENTAL RESULTS AND DISCUSSION

*Introduction*

The algorithm is evaluated against known network metrics and impact of rushing attack on multicast in mobile ad hoc network scheme specific network metrics. Comparison is done with rushing attacker node place at near sender, near receiver and uniformly distribution.

Metrics for Evaluation: The known network metrics to be used for performance evaluation is packet delivery ratio.

*Simulation Results*

We run several simulations under Linux, using the network simulator NS2 version ns-allinone-2.26. The simulation environment is composed of:

- area: 500*500 meters.
- number of nodes 50 - 100.
- simulation duration: 1000s.
- physical/Mac layer: IEEE 802.11 at 2Mbps, 250 meters transmission range.
- mobility model: random waypoint model with no pause time, and mode
- movement speed 0m/s, 1m/s and 10m/s.
- Using routing protocols are AODV and MAODV under NS2.26.

### A. Rushing attack at near sender (One sender and 5 receivers)

When the rushing attack happens at near sender in ad hoc network, the attack success rate is average because it has to search only the intermediate node. If there is no rushing attack in the network then the average attack success rate will be least.

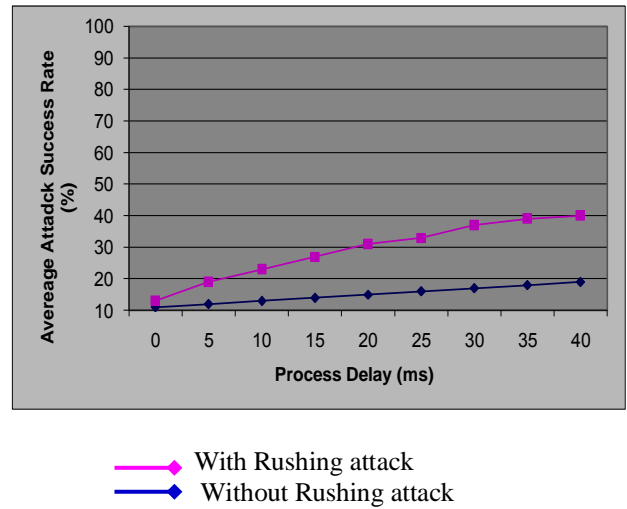

With Rushing attack
Without Rushing attack

Figure 7 Rushing attack at near sender

### B. Rushing Attack at Near Receiver

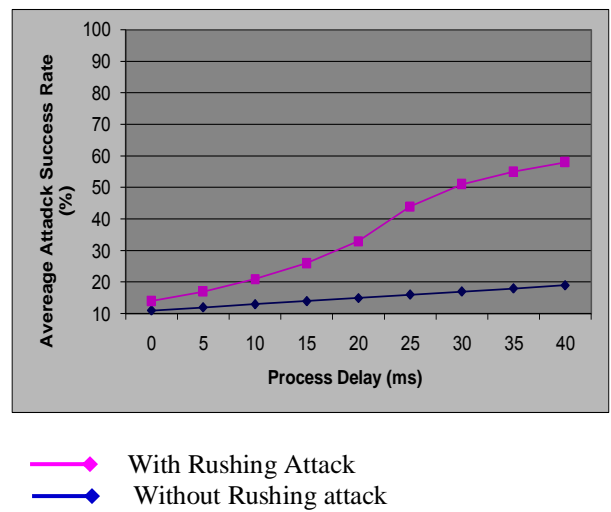

With Rushing Attack
Without Rushing attack

Figure 8 Rushing attack at near receiver

The figure 8 shows that the Attack Success Rate goes high, because of Rushing node is placed near receiver, because most of the forward node will contain all the packets. Since the attacker node is near to the receiver, it can gets the packet when the packet reaches the forward node near the receiver. Therefore, the receiver node get the packet quickly from the near attacker node and the impact of attack is highly harmful.

### C. Rushing Attack at Anywhere

The figure 9 shows that the Attack Success Rate goes least rate, because of Rushing node is placed anywhere. The attacker node is not placed at near sender or near receiver. The rushing node is placed anywhere (i.e. forward node in group). The forwarded node (Rushing attacker) taps the packet and quickly forwards the packets to the next node. Therefore, the chance of getting the packet from the attacker node depends on the upcoming nodes and so the impact of attack is least when its compare to near receiver's Attack Success Rate, is slightly higher than the near sender in which the Attack Success Rate is low

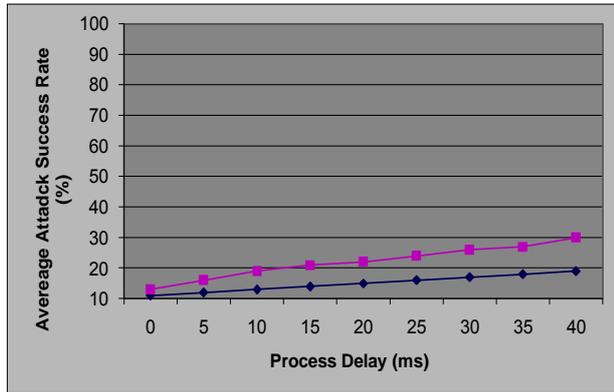

With rushing attack

Without rushing attack

Figure 9 Rushing Attack at anywhere

## V. CONCLUSION AND FUTURE DIRECTIONS

### A. Conclusion

The Rushing attacks are more likely to succeed in a multicast session where the number of multicast senders is small and/or the number of multicast receivers is large. The goal of the project is to draw the graph based on the rushing attack position in the network. With respect to the attack positions, the best position to launch rushing attacks is at the near receiver, have the highest success rates. The rushing attack near sender have the low success rate and final attack position is likely to take place anywhere in the network, have the least success rate.

### B. Future Directions

- In this project deals with one sender and multiple receivers in multicast ad hoc network. Apart from this there are chances to enhance it to have multiple senders and multiple receivers in multicast ad hoc network.

- In this project , it is assumed to have only one attacker node in the network for future it can be extended by adding more attacker nodes in the network.

## REFERENCES

[1]  Ping Yi, Zhoulin Dai, Shiyong Zhang, Yiping Zhong, "A New Routing Attack in Mobile Ad Hoc Networks International Journal of Information Technology Vol. 11 No. 2, pages 83 – 94.

[2]  Bruschi, D. and Rosti, E., "Secure Multicast in Wireless Networks of Mobile Hosts: Protocols and Issues", Mobile Networks and Applications, Volume 7, 2002, pp 503 - 511.

[3]  Moyer, M.J., Rao, J.R. and Rohatgi, P., "A Survey of Security Issues in Multicast Communication",IEEE Network, Nov.-Dec. 1999, pp. 12 – 23.

[4]  Dr. Jiejun Kong, " $GVG - RP$: A Net-centric Negligibility-based Security Model for Self-organizing Networks".

[5]  S.Corson,    J.Macker,    "Mobile    ad    hoc Networking(MANET):Routing Protocol Performance Issues and Evaluation Considerations, RFC 2501, January 1999.

[6]  C. Schuba, I. Krsul, M. Kuhn, E. Spafford, A. Sundaram, D. Zamboni, Analysis of a Denial of Service Attack on TCP, Proceedings of the 1997 IEEE Symposium on Security and Privacy.

[7]  Haining Wang, Danlu Zhang, and Kang G. Shin, Detecting    SYN    Flooding    Attacks,    IEEE INFOCOM'2002, New York City, 2002

[8]  Jiejun Kong, Xiaoyan Hong, Mario Gerla, " A new set of passive routing attacks in mobile ad hoc networks ",This work is funded by MINUTEMAN project and related STTR project of Office of Naval Research Pages 1- 6.

[9]  Jiejun Kong, Xiaoyan Hong, Mario Gerla, " Modeling Ad-hoc Rushing Attack in a Negligibility-based Security Framework", September 29, 2006, Los Angeles, California, USA.

[10]  Hoang Lan Nguyen , Uyen Trang Nguyen, "A study of different types of attacks on multicast in mobile ad hoc networks" Ad Hoc Networks 6 (2008) pages 32– 46.

[11]  S.J. Lee, W. Su, M. Gerla, " On-Demand Multicast Routing Protocol in Multihop Wireless Mobile Networks ", ACM/ Kluwer Mobile Networks and Applications 7 (6) (2002) 441– 453.

[12]  Imad Aad,   Jean-Pierre Hubaux ,   Edward W. Knightly, " Impact of Denial of Service Attacks on Ad Hoc Networks "

[13]  Y.-C. Hu, A. Perrig, and D. B. Johnson, "Ariadne: A secure ondemand routing protocol for ad hoc networks," in *Proceedings MobiCom 2002*, September 2002.

[14]  M. Zapata and N. Asokan, "Securing ad hoc routing protocols," in *Proceedings of the ACM Workshop on Wireless Security (WiSe)*, 2002.

[15]  Y.-C. Hu, A. Perrig, and D. B. Johnson, "Efficient security mechanisms for routing protocols," in *Network and Distributed System Security Symposium, NDSS*, 2003.

[16]  YihChun Hu,   Adrian Perrig, David B. Johnson, " Rushing Attacks and Defense in Wireless Ad Hoc Network  Routing Protocols " , WiSe 2003, September 19, 2003, San Diego California, USA Copyright 2003 ACM.

[17]  Yang, H., Luo, H., Ye, F., Lu, S., and Zhang, L., "Security in Mobile Ad Hoc Networks: Challenges and Solutions", IEEE Wireless Communications, Volume 11, Issue 1, February 2004, pp. 38 – 47.

[18]  Besemann,   C.,   Kawamura,   S.   and   Rizzo,   F., "Intrusion    Detection  System  in  Wireless  Ad-Hoc Networks: Sybil Attack Detection and Others".